\documentclass{article}

\usepackage{multirow}
\usepackage[hyphens]{url}
\usepackage{hyperref}
\usepackage{algorithm}
\usepackage{algpseudocode}
\usepackage{amsmath}
\usepackage{array}
\usepackage{varwidth}
\usepackage{caption}
\usepackage{numprint}
\usepackage{graphicx}
\usepackage{acronym}

\usepackage{floatpag}
\floatpagestyle{empty}

\newcommand{\defword}[1]{\textbf{\boldmath{#1}}}

\begin{document}

\npthousandsep{ }

\title{\Large\bf Solving Large Imperfect Information Games Using CFR$^+$}

\author{Oskari Tammelin\\ot@iki.fi}

\maketitle

\begin{abstract}

Counterfactual Regret Minimization and variants (e.g. Public Chance Sampling CFR and Pure CFR) have been known as the best approaches for creating approximate Nash equilibrium solutions for imperfect information games such as poker. This paper introduces CFR$^+$, a new algorithm that typically outperforms the previously known algorithms by an order of magnitude or more in terms of computation time while also potentially requiring less memory.
\end{abstract}

\section{Introduction}
\label{sec:introduction}

Counterfactual regret minimization (CFR)\cite{zinkevich2007regret} is a technique for solving extensive-form games with imperfect information. CFR and its numerous variants are based on \defword{regret-matching}\cite{hart2000simple}, a self-playing game-theoretic technique that chooses strategies proportional to positive cumulative regrets. In this technical report, we introduce a new CFR-like algorithm that instead uses a new variant of regret-matching that is empirically shown to perform better by an order magnitude or more.

\section{Algorithm}
\label{sec:algorithm}

The most straightforward CFR variant is often called "vanilla CFR" (or just CFR). In this variant, the entire tree is traversed at each iteration and no sampling is performed. There are four common ways vanilla CFR can be implemented in:

\vspace{0.2cm}
\begin{tabular}{|c|c|}
\hline 
Simultaneous updates / scalar-form & Alternating updates / scalar-form \\ 
\hline 
Simultaneous updates / vector-form & Alternating updates / vector-form \\ 
\hline 
\end{tabular} 
\vspace{0.2cm}

With \textit {Simultaneous updates}, each iteration has a single pass where both players' regrets and strategies are updated. In \textit{alternating updating}, there are two passes, in which one player is updated in each pass. The difference between scalar and vector forms can be found in \cite{johanson2012efficient}. CFR$^+$ is a \textit{vector-form, alternatingly updating} algorithm.

CFR$^+$ substitutes the regret-matching algorithm used in CFR with a new algorithm, \defword{regret-matching$^+$}. Using the definition of \defword{counterfactual value} $v_i(\sigma, I)$ in \cite{johanson2012efficient}, define the \defword{cumulative counterfactual regret$^+$} at information set $I$
for action $a$ up to time $T$ as

\begin{equation*}
R^{+,T}_i(I,a) =
\begin{cases}
\max\{v_i(\sigma^T_{I\rightarrow a},I)-v_i(\sigma^T,I), 0\}	& \text {$T = 1$} \\
\max\{R^{+,T-1}_i(I,a) + v_i(\sigma^T_{I\rightarrow a},I)-v_i(\sigma^T,I), 0\}	& \text {$T > 1$} \\
\end{cases}
\end{equation*}
The new strategy is produced by:
\begin{equation*}
\sigma^{T+1} = \begin{cases}
\frac{R^{+,T}_i(I,a)}{\sum_{a^\prime \in A(I)}R^{+,T}_i(I,a^\prime)} & \text { if the denominator is positive } \\
\frac{1}{|A(I)|} & \text { otherwise } \\
\end{cases}
\end{equation*}
Unlike in CFR, with CFR$^+$ the current strategy profile empirically either "almost" converges or converges to an approximate Nash equilibrium directly, so no averaging step is necessary. However, using \defword{weighted averaging}, the strategies can be further optimized to converge faster. The weight sequence used in CFR$^+$ is simply $w^T = \max{\{T-d, 0\}}$, where $d$ is the averaging delay in number of iterations. Figure \ref{fig:nlhe} shows an example of the effect delaying can have. 

The implementation is described in more detail in Algorithm \ref{alg:cfrplus}. See\cite{johanson2012efficient} for the meaning of the symbols used.

\section{Compressibility}
\label{sec:compressibility}

Another advantage of CFR$^+$ is that many of the cumulative regret values are zero, whereas in CFR, negative regret continues to accumulate indefinitely. This reduces the \defword{entropy} of the data needed during the computation. Using techniques such as fixed-point arithmetic (instead of floating-point arithmetic), linear and non-linear prediction, context modeling and arithmetic coding (or Asymmetric Numeral Systems), high compression ratios can be attained in large poker games. Details of CFR$^+$ and compression are outside the scope of this paper and will be discussed in a future publication.

\section{Results}
\label{sec:results}

We compare CFR$^+$ to the fastest CFR variant seen during our tests in one-card poker, the vector-form vanilla CFR with alternating regret updates --- the same form CFR+ uses. 

\begin{figure}
\advance\leftskip-2cm
\includegraphics[width=1.5\textwidth]{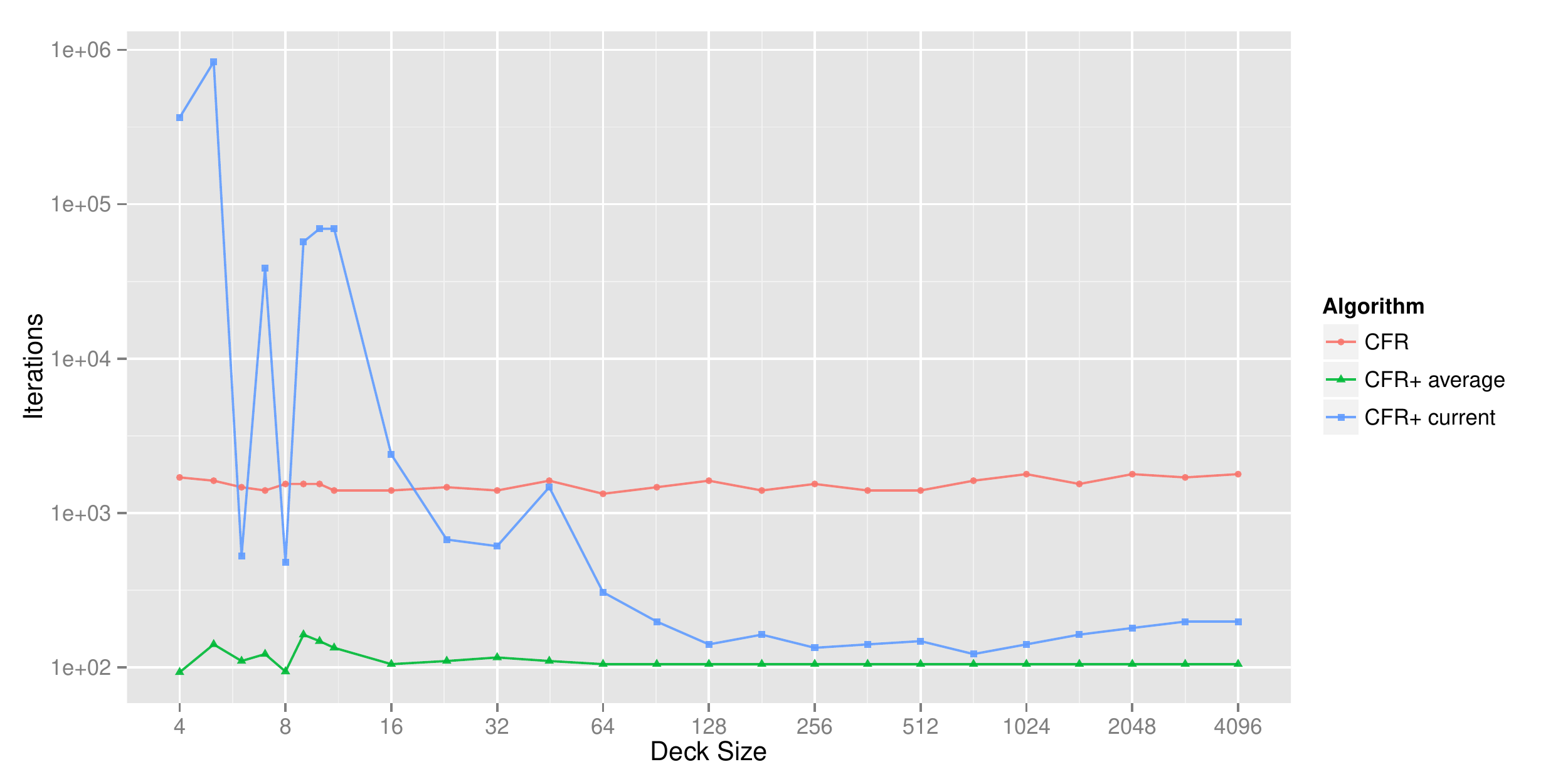}
\caption{One-card poker iterations vs. deck size}
\label{fig:ocp}

\includegraphics[width=1.5\textwidth]{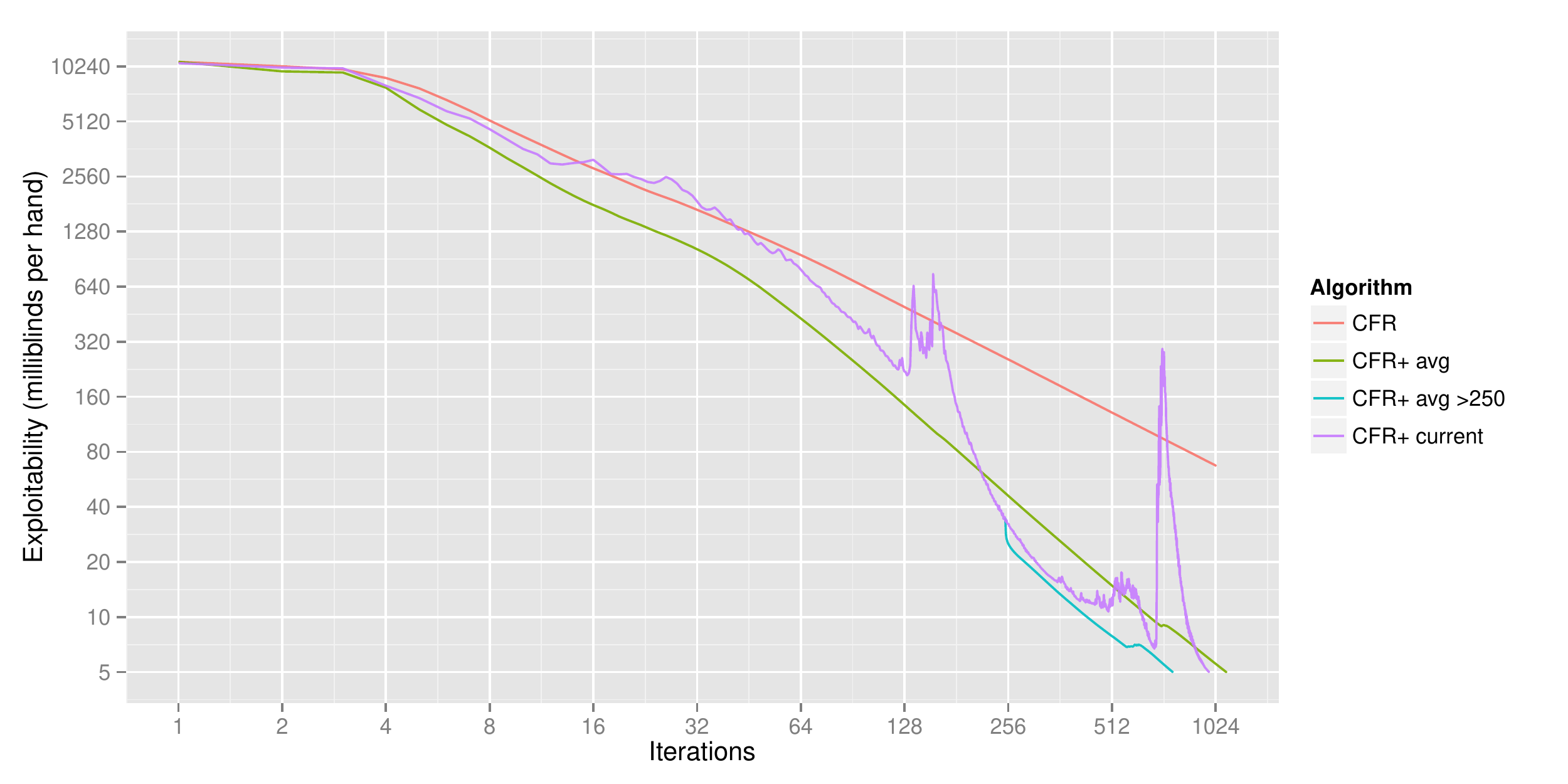}
\caption{No Limit Texas Hold'em flop subgame}
\label{fig:nlhe}
\end{figure}

Figure \ref{fig:ocp} shows a comparison of CFR$^+$ with vanilla CFR in a one-card poker game using various deck sizes. The vertical axis is the number of iterations needed to get the exploitability below one millibets per hand. We observe that (1) CFR$^+$ current strategy seems to converge slowly when the deck size is small and (2) CFR$^+$ average strategy converges more than an order of magnitude faster than CFR.

Figure \ref{fig:nlhe} shows a No Limit Texas Hold'em subgame starting after the preflop betting, with a 5-milliblind target exploitability. An action abstraction that allows three different bet sizes is used. Again, we see a similar difference in performance.

\begin{algorithm}
\caption{CFR$^+$ Algorithm}
\begin{algorithmic}[1]

\State Initialize regret tables: $\forall{I}, r_I[a]\gets 0$.
\State Initialize cumulative strategy tables: $\forall{I}, s_I[a]\gets 0$.
\State

\Function{CFR$^+$}{$h, i, w, \vec{\pi}_{-i}$}:
	\If {$h \in Z$}
		\State \textbf{return} $\vec{f}_{c,i}(h)\odot\vec{u}_i\left(h\mid\vec{\pi}_{-i}\odot\vec{f}_{c,-i}(h)\right)$
	\EndIf

	\State $\vec{u}\gets\vec{0}$
	
	\If {$h \in \mathcal{P}$}
		\For {$a \in A(h)$}
			\State {$\vec{u}^\prime\gets $CFR$^+(ha, i, w, \vec{\pi}_{-i}$)}
			\State {$\vec{u}\gets\vec{u}+f_c(a|h)\vec{u}^\prime$}
			
		\EndFor
		
   \State \textbf{return} $\vec{u}$
	\EndIf

	\State $\vec{I}\gets$ lookupInfosets($h$)
	\State $\vec{\sigma}\gets$ regretMatching$^+(\vec{I}$)

	\If {$P(h) = i$}
		\For {$a \in A(h)$}
			\State {$\vec{u}^\prime\gets $CFR$^+(ha, i, w, \vec{\pi}_{-i}$)}
			\State {$\vec{m}[a]\gets\vec{u}^\prime$}
			\State {$\vec{u}\gets\vec{u}+\vec{\sigma}[a]\odot\vec{u}^\prime$}
		\EndFor
		\For {$I\in\vec{I}$}
			\For {$a \in A(I)$}
				\State {$r_I[a]\gets \max\{r_I[a] + m[a][I] - u[I], 0\} $}
			\EndFor
		\EndFor
	\Else
		\For {$a \in A(h)$}
			\State {$\vec{\pi}^\prime_{-i}\gets\vec{\sigma}[a]\odot\vec{\pi}_{-i}$}
			\State {$\vec{u}^\prime\gets $CFR$^+(ha, i, w, \vec{\pi}^\prime_{-i}$)}
			\State {$\vec{u}\gets\vec{u}+\vec{u}^\prime$}
		\EndFor
		\For {$I\in\vec{I}$}
			\For {$a \in A(I)$}
				\State {$s_I[a]\gets s_I[a] + {\pi}_{-i}[I] {\sigma}[a][I]w $}
			\EndFor
		\EndFor
	\EndIf

   \State \textbf{return} $\vec{u}$
\EndFunction
\State
\Function{CFR$^+$}{d}:
	\For {$t \in \left\{1, 2, 3, \ldots\right\}$}
		\For {$i \in N$}
			\State CFR$^+(\emptyset, i, \max\{t-d,0\}, \vec{1}$)
		\EndFor

	\EndFor

\EndFunction

\end{algorithmic}
\label{alg:cfrplus}
\end{algorithm}

\section{Conclusion}
\label{sec:conclusion}

In this technical report, we briefly introduced CFR$^+$, compared it to CFR, showing more than an order of magnitude improvement in converge time. In a future paper, we will attempt to explain the mathematics behind the algorithm in more detail.

\renewcommand{\abstractname}{Acknowledgements}
\begin{abstract}
Thanks to Ronan O'Ceallaigh and Michael Bowling for their help and advice.
\end{abstract}

\clearpage
\bibliographystyle{unsrt}
 \bibliography{cfrp}

\end{document}